\documentclass[floatfix,twocolumn,aps,prd,showpacs,amsmath,amssymb]{revtex4}
\usepackage{epsfig}
\usepackage{graphicx}
\usepackage{subfigure}
\usepackage[latin1]{inputenc}

\begin{document}

\title{Quantum radiation reaction force on a one-dimensional cavity with two relativistic moving mirrors}

\author{Danilo T. Alves$^{1}$, Edney R. Granhen$^{2}$ and Wagner P. Pires$^{3}$}
\affiliation{(1) - Faculdade de F\'\i sica, Universidade Federal do
Par\'a, 66075-110, Bel\'em, PA,  Brazil\\(2) - Centro Brasileiro de Pesquisas Físicas, Rua Dr. Xavier Sigaud,
150, 22290-180, Rio de Janeiro, RJ, Brazil\\(3)
Instituto de Física, Universidade Federal do Rio de Janeiro, 21941-972, Rio de Janeiro, Brazil}
\date{\today}

\begin{abstract}
We consider a real massless scalar field inside a cavity with two moving mirrors 
in a two-dimensional spacetime,
satisfying Dirichlet boundary condition at the instantaneous position of the boundaries,
for arbitrary and relativistic laws of motion.
Considering vacuum as the initial field state, we 
obtain formulas for the exact value of the energy density of the field and the quantum force acting on the boundaries,
which extend results found in previous papers \cite{Li-Li-PLA-2002,Li-Li-APS-2003,alves-granhen-silva-lima-2010,Cole-Schieve-2001}. 
For the particular cases of a cavity with just one moving boundary, 
non-relativistic velocities, or in the limit of infinity length of the cavity
(a single mirror), our results coincide with those found in the
literature.
\end{abstract}


\maketitle

The Dynamical Casimir effect has been investigated since the 1970s
\cite{Moore-1970, Fulling-Davies-PRS-1976-I,trabalhos-pioneiros},
and has attracted growing attention. It is related to problems like particle creation in cosmological models 
and radiation emitted by collapsing black holes \cite{Fulling-Davies-PRS-1976-I,Birrel-Davies-1982},
decoherence \cite{descoerencia}, entanglement \cite{emaranhamento}, 
the Unruh effect \cite{Luis-RMP-2008}, among others.
Models of a single mirror have been investigated 
and
also cavities with one moving boundary have been studied in many papers
(for a review see Ref. \cite{Dodonov-Revisao-2001}).
In contrast, the problem of a cavity with two moving boundaries
has been investigated recently and
relatively few papers on this subject are found in the literature 
(for instance, Refs. \cite{Reynaud-Lambrecht-Jaekel-1996,Ji-Jung-Soh-1998,duas-fronteiras,Machado-Maia-Neto-PRD-2002,Dalvit-PRA-1999}).
A cavity with two oscillating mirrors can exhibit situations of constructive and destructive interference
in the number of created particles, 
depending on the relation among the phase difference of each boundary,
the amplitudes and frequencies of oscillation
\cite{Reynaud-Lambrecht-Jaekel-1996,Ji-Jung-Soh-1998,Dalvit-PRA-1999}.
Ji, Jung and Soh \cite{Ji-Jung-Soh-1998}, considering
the expansion of the quantizing field over a instantaneous basis and a perturbative approach,
investigated the problem of interference in the particle creation for 
a one-dimensional cavity with two boundaries moving according to prescribed, non-relativistic and oscillatory (small amplitudes) 
laws of motion. 
Dalvit and Mazzitelli \cite{Dalvit-PRA-1999} 
extended the field solution
obtained by Moore \cite{Moore-1970} for the case of
a one-dimensional cavity with two moving boundaries,
deriving a set of generalized Moore´s equations, also obtaining the expected energy-momentum tensor for this model, generalizing
the corresponding formula obtained by Fulling and Davies \cite{Fulling-Davies-PRS-1976-I}.
In Ref. \cite{Dalvit-PRA-1999} the set of generalized Moore´s equations
was solved for the case of a resonant
oscillatory movement with small amplitude,
via renormalization-group procedure.
Li and Li \cite{Li-Li-PLA-2002} applied the geometrical method, proposed by Cole and Schieve \cite{Cole-Schieve-1995},
to solve exactly the generalized Moore equations obtained by Dalvit and Mazzitelli \cite{Dalvit-PRA-1999},
and also used numerical methods to obtain the behavior of the energy density in a cavity for particular sinusoidal laws of motion, with small amplitude \cite{Li-Li-APS-2003}. 
On the other hand, as far as we know, there is no paper in literature devoted to 
obtain formulas which enable us to get directly exact values 
for the quantum force and energy density in a nonstatic cavity for 
arbitrary laws of motion for the moving boundaries, 
including non-oscillating movements with large amplitudes, which are out of reach of the perturbative approaches found in the literature. 

In the present paper we consider a real massless scalar field
satisfying the Klein-Gordon equation (we assume throughout this paper $\hbar=c=1$):
$
\left(\partial _{t}^{2}-\partial _{x}^{2}\right) \phi \left(
t,x\right) =0,
$
and obeying Dirichlet conditions imposed at the left boundary located at $x=L(t)$,
and also at the right boundary located at $x=R(t)$, where $L(t)$ and  $R(t)$ are arbitrary prescribed laws
of motion, with $R(t<0)=L_0$ and $L(t<0)=0$,
where $L_0$ is the length of the cavity in the static situation.
Considering vacuum as the initial field state, we 
obtain formulas for the exact value of the energy density of the field and the quantum force acting on the boundaries,
showing that the energy density in a given point of the spacetime can be obtained by tracing back a sequence of null lines, connecting the value of the energy density at the given spacetime point to a certain known value of the energy at a point in the ``static zone'', where the initial field modes are not affected by the disturbance caused by the movement of the boundaries.
Our formulas generalize those found in literature \cite{alves-granhen-silva-lima-2010}, 
where this problem is approached for a cavity
with only one moving mirror. For the particular cases of a cavity with just one moving boundary, 
non-relativistic velocities, or in the limit of large length of the cavity
(a single mirror), our results coincide with those found in the literature \cite{Cole-Schieve-2001,
alves-granhen-lima-2008,alves-granhen-silva-lima-2010}.
 
Let us start considering the field operator, solution of the wave equation, given by \cite{Dalvit-PRA-1999}
\begin{equation*}
\hat{\phi} \left( t,x\right) =\sum_{k=1}^{\infty }\left[ \hat{a}_{k}\psi _{k}\left(t,x\right) +\hat{a}_{k}^{\dag }\psi _{k}^{\ast }\left( t,x\right) \right], 
\label{campo}
\end{equation*}
where the field modes are 
\begin{equation}
\psi _{k}\left(t,x\right) =\frac{i}{\sqrt{4\pi k}}\left[ e^{-ik\pi G\left(v\right) }-e^{-ik\pi F\left(u\right) }\right],
\label{psik} 
\end{equation}
with $v=t+x$, $u=t-x$, and 
\begin{subequations}
\label{LG}
\begin{eqnarray}
G\left[ t+L\left( t\right) \right] -F\left[ t-L\left( t\right) \right]  &=&0
\label{para L} \\
G\left[ t+R\left( t\right) \right] -F\left[ t-R\left( t\right) \right]  &=&2.
\label{para R}
\end{eqnarray}
\end{subequations}
The set of equations (\ref{LG}), obtained by Dalvit and Mazzitelli exploiting
the conformal invariance of the model \cite{Dalvit-PRA-1999}, 
is a generalization of the Moore equation \cite{Moore-1970}, which can be recovered doing $L(t)=0$ in these equations. 
The renormalized energy density in the cavity is given by \cite{Dalvit-PRA-1999}
\begin{equation}
\left\langle T_{00}\left( t,x\right) \right\rangle=-f_{G}\left( v\right) -f_{F}\left( u\right),
\label{tensor}
\end{equation}
where
\begin{subequations}
\begin{eqnarray}
f_{G}(z) &=&\frac{1}{24\pi }\left\{ \frac{G^{\prime \prime \prime }(z)}{G^{\prime }(z)}%
-\frac{3}{2}\left[ \frac{G^{\prime \prime }(z)}{G^{\prime }(z)}\right] ^{2}+\frac{%
\pi ^{2}}{2}[G^{\prime}(z)]^2\right\},\nonumber\\
&&  \label{f de G} \\
f_{F}(z) &=&\frac{1}{24\pi }\left\{ \frac{F^{\prime \prime \prime }(z)}{F^{\prime }(z)}%
-\frac{3}{2}\left[ \frac{F^{\prime \prime }(z)}{F^{\prime }(z)}\right] ^{2}+
\frac{\pi ^{2}}{2}[F^{\prime}(z)]^2\right\}. \nonumber\\
&& \label{f de F}
\end{eqnarray}
\end{subequations}

In the present paper we use Eqs. (\ref{para L}), (\ref{para R}), (\ref{f de G}) and (\ref{f de F}) 
to obtain the following set of equations for the functions $f_{G}$ and $f_{F}$:
\begin{subequations}
\label{cincos}
\begin{eqnarray}
f_{G}\left[ t+R\left( t\right) \right]  &=&f_{F}\left[ t-R\left( t\right)\right] A_{R}\left( t\right) +B_{R}\left( t\right), \nonumber \\ &&  \label{fG de t + q} \label{fGdeR} \\
f_{G}\left[ t+L\left( t\right) \right]  &=&f_{F}\left[ t-L\left( t\right)\right] A_{L}\left( t\right) +B_{L}\left( t\right), \nonumber \\ && \label{fG de t + L 2} \label{fGdeL} 
\end{eqnarray}
\end{subequations}
with
\begin{equation}
A_{q}\left(t\right)=\left[ \frac{1-q^{\prime }\left( t\right) }{%
1+q^{\prime }\left( t\right) }\right]^{2},
\label{A}
\end{equation}
\begin{eqnarray}
B_{q}\left( t\right)  &=&-\frac{1}{12\pi }\frac{q^{\prime \prime \prime
}\left( t\right) }{\left[ 1+q^{\prime }\left( t\right) \right] ^{3}\left[
1-q^{\prime }\left( t\right) \right] }  \nonumber \\
&&-\frac{1}{4\pi }\frac{q^{\prime \prime 2}\left( t\right) q^{\prime }\left(
t\right) }{\left[ 1+q^{\prime }\left( t\right) \right] ^{4}\left[
1-q^{\prime }\left( t\right) \right] ^{2}}, \label{B}
\end{eqnarray}
where, hereafter, $q$ can represent $R$ or $L$.
\begin{figure}
\centering
\includegraphics[scale=.6]{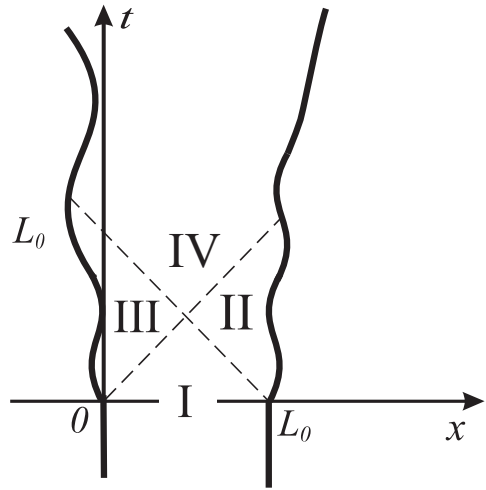}
\caption{Boundaries trajectories (solid lines). The dashed lines are null lines separating region I from II and III, and these ones from region IV.}
\label{static-zones}
\end{figure}
Eqs. (\ref{fGdeR}) and (\ref{fGdeL}) are an extension of the corresponding
equation for $f$, valid for a cavity with just one
moving mirror, found in Ref. \cite{Cole-Schieve-2001}.
If we consider the particular case of $L(t)=0$ in Eq. (\ref{cincos}), 
we recover the corresponding result found in Ref. \cite{Cole-Schieve-2001}.
For ($t<0$) we have $f_{G}(v)=f_{F}(u)=f^{(s)}=\pi/(48L_0^2)$, and $\left\langle T_{00}\right\rangle=-\pi/(24L_0^2)$,
which is the Casimir energy density for this model. 

Now, our aim is to solve the Eqs. (\ref{fGdeR}) and (\ref{fGdeL}) recursively, using a geometrical point of view.
Let us examine the cavity in the nonstatic situation ($t > 0$). The field modes in Eq. (\ref{psik}) are formed by left and
right-propagating parts. As causality requires, the field in region I ($v < L_0$ and $u < 0$) (see Fig. \ref{static-zones}) is not affected by the
boundaries motion, so that, in this sense, this region is considered as a ``static zone''. In region II ($v>L_0$ and $u < 0$), the right-propagating parts of the field modes remain unaffected by the boundaries motion, so that region II is also a static zone for these modes. On the other hand, the left-propagating parts in region II are, in general, affected by the boundary movement. Similarly, in region III ($u>0$ and $v<L_0$), the left-propagating parts of the field modes are not affected by the boundaries motion, but the right-propagating parts are. In region IV ($v>L_0$ and $u>0$), both the left and right-propagating parts are affected. In summary, the functions corresponding to the left and right-propagating parts of the field modes are considered in the static zone if their arguments
are, respectively $v< L_0$ and $u<0$. Then, we have $f_G(v<L_0)=f^{(s)}$ and $f_G(u<0)=f^{(s)}$.

\begin{figure}
\subfigure[\label{reflections2FnG1}  $n_G=1$]{\includegraphics[scale=.46]{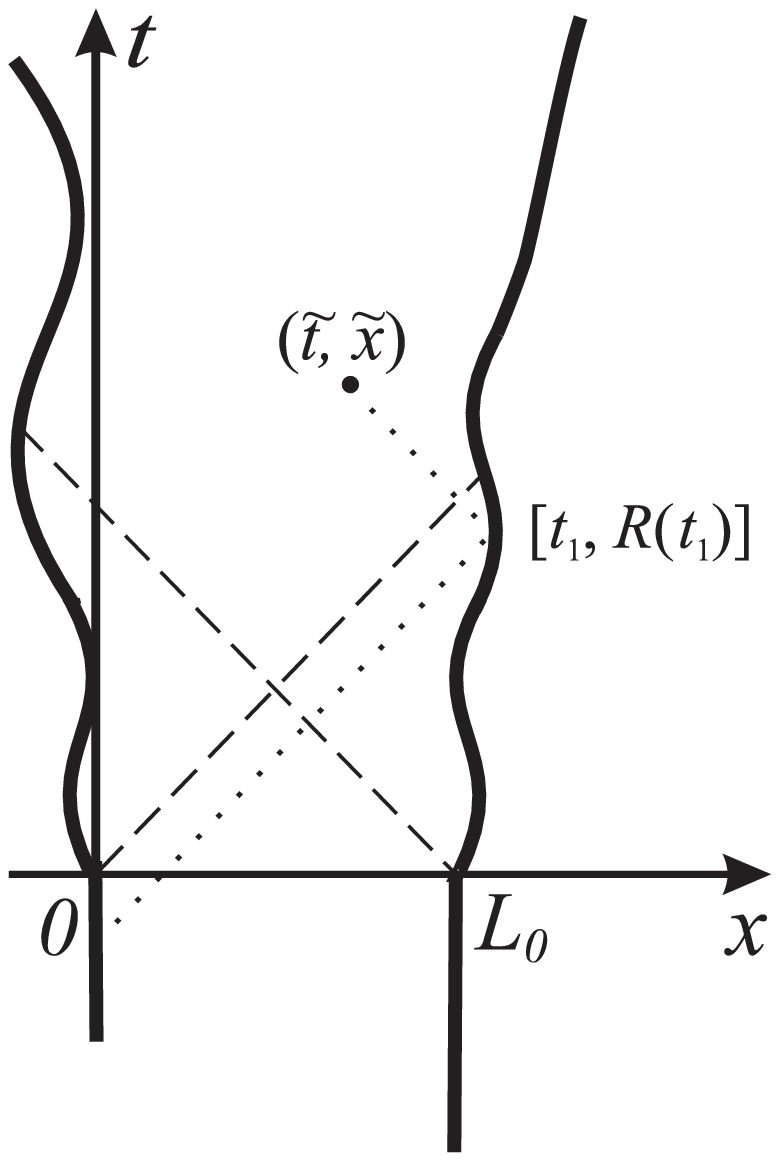}}
\subfigure[\label{reflections2F}  $n_G=2$]{\includegraphics[scale=.46]{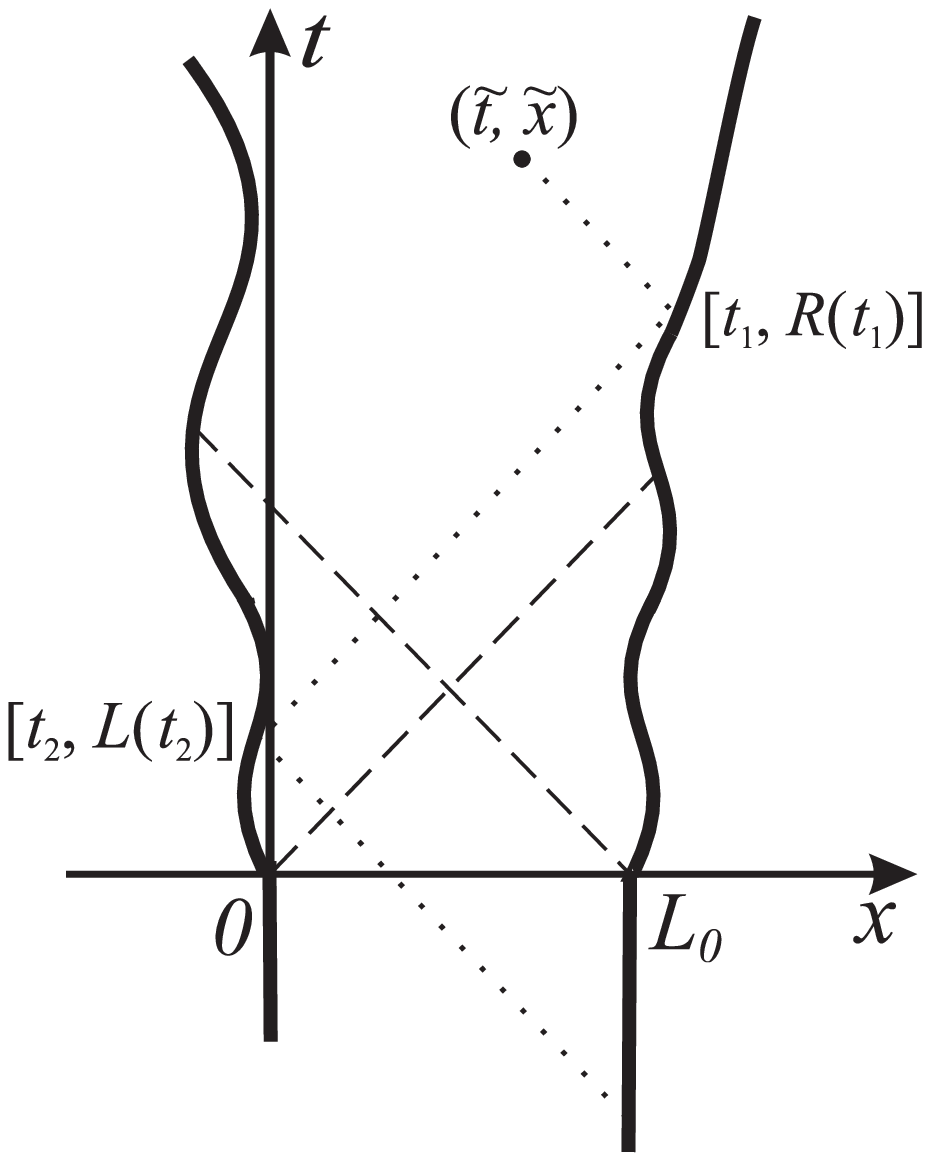}}
\caption{Sequence of null lines (dotted lines) connecting a point $(\tilde{t},\tilde{x})$ to a static zone.
The dashed lines are null lines separating region I from II and III, and these ones from region IV,
as presented in Fig. \ref{static-zones}. In Fig. \ref{reflections2FnG1}, we see the case of one reflection ($n_G=1$),
whereas in Fig. \ref{reflections2F} we see the case $n_G=2$.}
\end{figure}

For a certain spacetime point $\left(\tilde{t},\tilde{x}\right)$, the energy tensor $\left\langle T_{00}\left(\tilde{t},\tilde{x}\right)\right\rangle$ is known if its left and right-propagating parts, taken over, respectively, the null lines $v = z_1$ and $u = z_2$ (where $z_1 = \tilde{t} + \tilde{x}$ and $z_2 = \tilde{t} - \tilde{x}$), are known; or, in other words, $\left\langle T_{00}\left(\tilde{t},\tilde{x}\right)\right\rangle$ is known if $\left.f_G\left(v\right)\right|_{v=z_1}$ and $\left.f_F\left(u\right)\right|_{u=z_2}$ are known.
Li and Li \cite{Li-Li-PLA-2002} used a recursive method \cite{Cole-Schieve-1995} to obtain the functions $G$ and $F$ for general laws of motion of the boundaries, tracing back a sequence of null lines until a null line gets into the static zone where the $G$ or $F$ functions are known. Here, we adopt this method to obtain $f_G$ and $f_F$, extending the work done in Ref. \cite{Li-Li-PLA-2002}.
Let us assume that $\left(\tilde{t},\tilde{x}\right)$ belongs to region IV, and that the null line $v = z_1$ intersects the moving mirror trajectory at the point $\left[t_1,R\left(t_1\right)\right]$ (see Fig. \ref{reflections2FnG1}),
so that $\tilde{t} + \tilde{x}= t_1+R\left(t_1\right)$.
%
%
We have $\left.f_G\left(v\right)\right|_{v=z_1}=f_G\left[t_1+R\left(t_1\right)\right]$. Using the Eq. (\ref{fGdeR}), we get $f_{G}\left[ t_1+R\left( t_1\right) \right] =f_{F}\left[ t_1-R\left( t_1\right)\right] A_{R}\left( t_1\right) +B_{R}\left( t_1\right)$. If $t_1-R\left(t_1\right) < 0$, then the null line $u = t_1 - R\left(t_1\right)$ is already in the static zone (Fig. \ref{reflections2FnG1}), so that we can write $f_{F}\left[ t_1-R\left( t_1\right)\right]=f^{(s)}$, and also $f_{G}\left[ t_1+R\left( t_1\right) \right]  =f^{(s)} A_{R}\left( t_1\right) +B_{R}\left( t_1\right)$, and we can say that the number of reflections $n_G$ to get into the static zone is,
in this case,  $n_G=1$.
On the other hand, if $t_1-R\left(t_1\right) > 0$ (case shown in Fig. \ref{reflections2F}) we can draw another null line $v = t_2 + L\left(t_2\right)$ intersecting the world line of the left boundary at the point $\left[t_2,L\left(t_2\right)\right]$,
with $t_1-R\left(t_1\right)=t_2-L\left(t_2\right)$. In this case we have, using (\ref{fGdeL}), $f_{G}\left[ t_1+R\left( t_1\right) \right]=\left\{f_{G}\left[ t_2+L\left( t_2\right) \right]-B_L\left(t_2\right)\right\}A_R\left(t_1\right)/A_L\left(t_2\right)+B_R\left(t_1\right)$. 
If $t_2+L\left( t_2\right) < L_0$ (see Fig. \ref{reflections2F}), then $f_{G}\left[ t_2+L\left( t_2\right)\right]=f^{(s)}$,
$f_{G}\left[ t_1+R\left( t_1\right) \right]=\left\{f^{(s)}-B_L\left(t_2\right)\right\}A_R\left(t_1\right)/A_L\left(t_2\right)+B_R\left(t_1\right)$ and $n_G=2$.
If $t_2+L\left( t_2\right) > L_0$, we assume that the null line $v=t_2+L\left( t_2\right)$ intersects the right boundary at the point $\left[t_3,R\left(t_3\right)\right]$, then $t_2+L\left( t_2\right)=t_3+R\left(t_3\right)$ and we get $u=t_3-R\left(t_3\right)$. 
We repeat this procedure up to a null line gets into a static zone, where the function $f_{F}$ or $f_{G}$ is known. 
In summary, we obtain for $f_G$:
\begin{equation}
f_{G}\left( z\right) =f^{(s)} \tilde{A}_{G}\left( z\right)+ \tilde{B}_{G}\left( z\right),  
\label{fGz}
\end{equation}
where, for $n_G\left(z\right)$ even, we have
\begin{subequations}
\label{ABG-par}
\begin{equation}
\tilde{A}_{G}\left( z\right) =\prod_{k=0}^{\frac{n_{G}\left( z\right) }{2}}\left[ \left( 1-\delta _{k,0}\right) \frac{A_{R}\left[ t_{2k-1}(z)\right] }{A_{L}\left[ t_{2k}(z)\right] }+\delta _{k,0}\right],
\label{atilnGpar}
\end{equation}
\begin{eqnarray}
\tilde{B}_{G}\left( z\right)  &=&\sum_{k=0}^{\frac{n_{G}\left( z\right) }{2}}\left\{ \left( 1-\delta _{k,0}\right) \left[ \frac{B_{R}\left[t_{2k-1}(z)\right] A_{L}\left[ t_{2k}(z)\right] }{A_{R}\left[ t_{2k-1}(z)\right] }\right. \right.   \nonumber \\
&&\left. -B_{L}\left[ t_{2k}(z)\right] \right]  \nonumber \\
&&\left. \times \prod_{j=0}^{k}\left[ \left( 1-\delta _{j,0}\right) \frac{A_{R}\left[ t_{2j-1}(z)\right] }{A_{L}\left[ t_{2j}(z)\right]}+\delta _{j,0}\right] \right\}, \nonumber \\
&&
\label{btilnGpar}
\end{eqnarray}
\end{subequations}
with $\delta$ symbolizing Kronecker's delta function.
For $n_G\left(z\right)$ odd we have
\begin{subequations}
\label{ABG-impar}
\begin{equation}
\tilde{A}_{G}\left( z\right) =\prod_{k=0}^{\frac{n_{G}\left( z\right) -1}{2}}\left[ \frac{A_{R}\left[ t_{2k+1}(z)\right] }{\left( 1-\delta _{k,0}\right)A_{L}\left[ t_{2k}(z)\right] +\delta _{k,0}}\right],
\label{atilnGimpar}
\end{equation}
\begin{eqnarray}
\tilde{B}_{G}\left( z\right)  &=&\sum_{k=0}^{\frac{n_{G}\left( z\right) -1}{2}}\left\{ \left[ B_{R}\left[ t_{2k+1}(z)\right] -\left( 1-\delta _{k,0}\right)B_{L}\left[ t_{2k}(z)\right] \right] \right. \nonumber \\
&&\times \left. \prod_{j=0}^{k}\left[ \left( 1-\delta _{j,0}\right) \frac{A_{R}\left[ t_{2j-1}(z)\right] }{A_{L}\left[ t_{2j}(z)\right] }+\delta _{j,0}\right] \right\}. 
\label{btilnGimpar}
\end{eqnarray}
\end{subequations}
Note that the number $n_G$ of reflections and the sequence of instants $t_1,...,t_{n_G}$ depend on 
the argument $z$. The set of instants mentioned in Eqs. (\ref{ABG-par}) and (\ref{ABG-impar}) are calculated
via \cite{Li-Li-PLA-2002}:
\begin{eqnarray}
z&=&t_1+R(t_1),
\nonumber
\\
t_{2l+1}-R(t_{2l+1})&=&t_{2l+2}-L(t_{2l+2}),
\label{t-G}
\\
t_{2l+2}+L(t_{2l+2})&=&t_{2l+3}+R(t_{2l+3}),
\nonumber
\\
l&=&0,1,2...
\nonumber
\end{eqnarray}

\begin{figure}
\subfigure[\label{reflections2FnF1}  $n_F=1$]{\includegraphics[scale=.45]{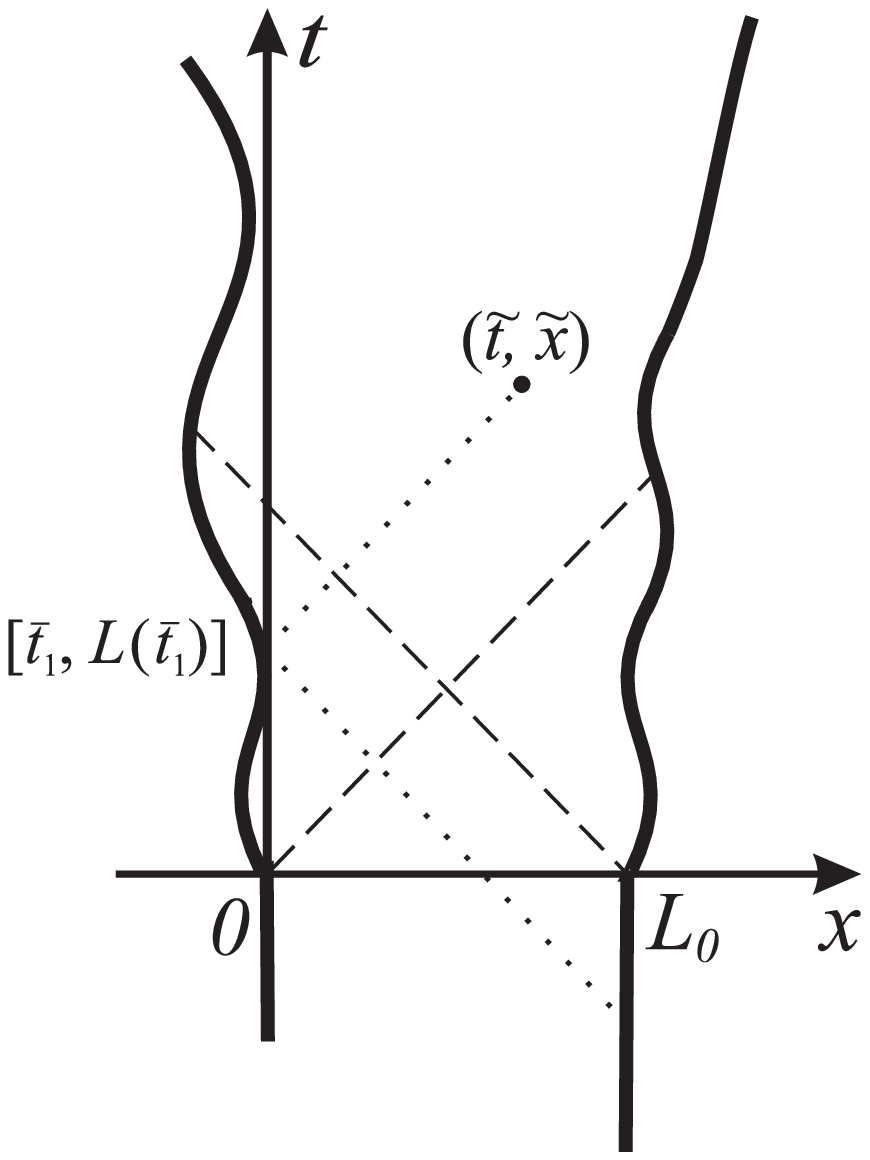}}
\subfigure[\label{reflections2FnF2}  $n_F=2$]{\includegraphics[scale=.45]{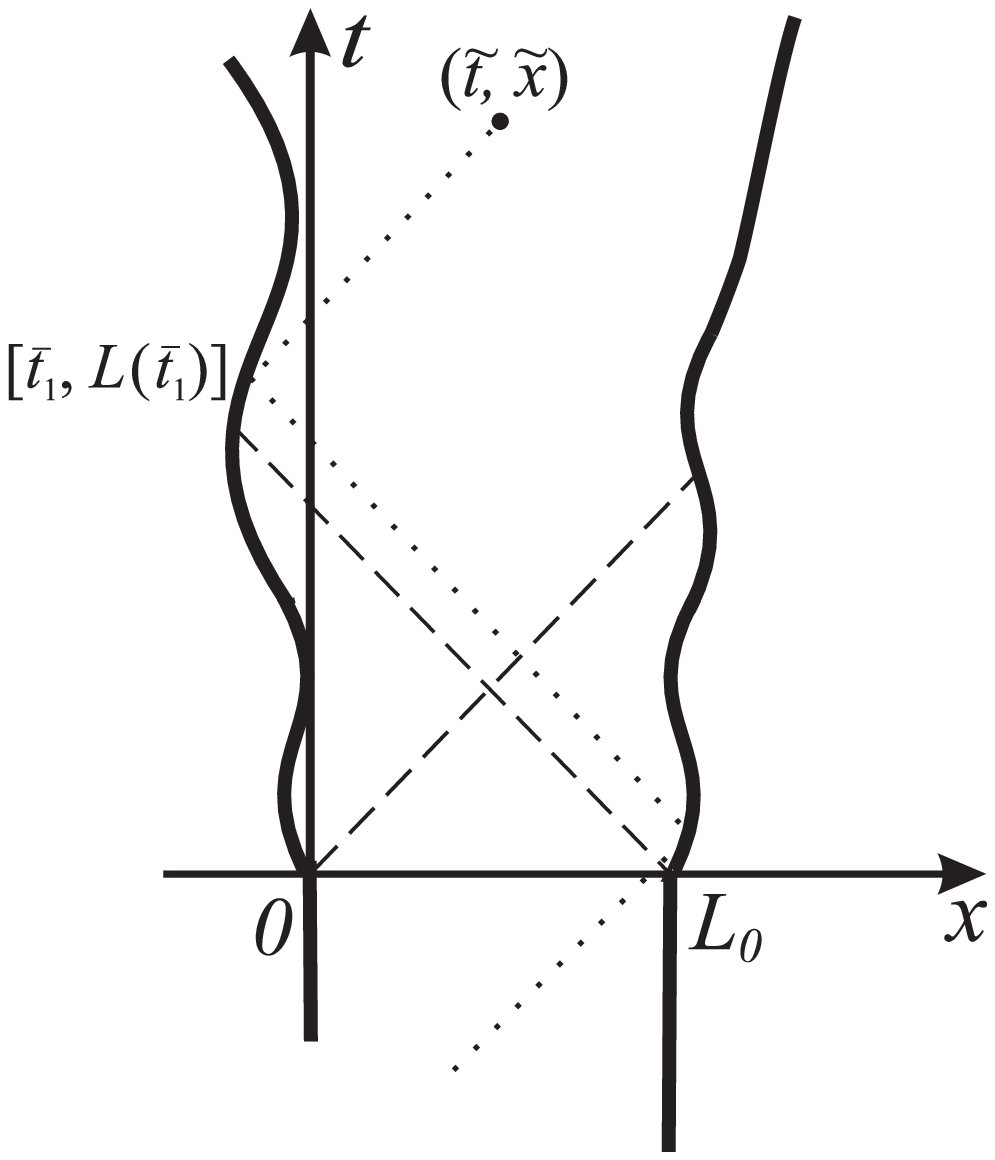}}
\caption{Sequence of null lines (dotted lines) connecting a point $(\tilde{t},\tilde{x})$ to a static zone.
The dashed lines are null lines separating region I from II and III, and these ones from region IV,
as presented in Fig. \ref{static-zones}. In Fig. \ref{reflections2FnF1}, we see the case of one reflection ($n_F=1$),
whereas in Fig. \ref{reflections2FnF2} we see the case $n_F=2$.}
\end{figure}

To solve recursively the set of equations (\ref{cincos}) for $f_{F}$,
we start assuming that the null line $u = \tilde{t} - \tilde{x}$ 
intersects the worldline of the left mirror at the point $\left[\bar{t}_1,L\left(\bar{t}_1\right)\right]$,
so that $\tilde{t} - \tilde{x}= \bar{t}_1-L\left(\bar{t}_1\right)$. 
Thus we have $\left.f_{F}\left(u\right)\right|_{u=z_2}=
f_F\left[\bar{t}_1-L\left(\bar{t}_1\right)\right]$.
Using the Eq. (\ref{fGdeL}), we get $f_{F}\left[ \bar{t}_1-L\left( \bar{t}_1\right) \right]
 =\left\{f_{G}\left[ \bar{t}_1+L\left( \bar{t}_1\right)\right] -B_{L}\left( \bar{t}_1\right)
\right\}/A_{L}\left( \bar{t}_1\right)$. If $\bar{t}_1+L\left(\bar{t}_1\right) 
< L_0$, then the null line $v = \bar{t}_1 + L\left(\bar{t}_1\right)$ is already 
in the static zone, so that we can write $f_{G}\left[ \bar{t}_1+L\left( \bar{t}_1\right)
\right]=f^{(s)}$, and also $f_{F}\left[ \bar{t}_1-L\left( \bar{t}_1\right) \right]
 =\left\{ f^{(s)}-B_{L}\left( \bar{t}_1\right)\right\}/A_{L}\left( \bar{t}_1\right)$,
 and we can say that the number of reflections $n_F$ to get into the static zone is, 
in this case,  $n_F=1$ (see Fig. \ref{reflections2FnF1}).
On the other hand, 
if $\bar{t}_1+L\left(\bar{t}_1\right) >
L_0$ (as shown in Fig. \ref{reflections2FnF2}) we 
need to find $f_{G}\left[ \bar{t}_1+L\left( \bar{t}_1\right)\right]$
recursively via Eq. (\ref{fGz}).
In general, we get
%
%
\begin{equation}
f_{F}\left( z\right) =f^{(s)} \tilde{A} _{F}\left( z\right)+\tilde{B} _{F}\left( z\right),
\label{fFz}
\end{equation}
where
\begin{subequations}
\label{ABF}
\begin{equation}
\tilde{A}_{F}\left( z\right) = \frac{\tilde{A}_{G}\left\{ \bar{t}_1(z)+L[\bar{t}_1(z)]\right\}}{A_L[\bar{t}_1(z)]},
\label{atilnFpar}
\end{equation}
\begin{equation}
\tilde{B}_{F}\left( z\right) = \frac{\tilde{B}_{G}\left\{ \bar{t}_1(z)+L[\bar{t}_1(z)]\right\}-B_L[\bar{t}_1(z)]}{A_L[\bar{t}_1(z)]},
\label{btilnFpar}
\end{equation}
\end{subequations}
with the function $\bar{t}_1(z)$ calculated via
\begin{eqnarray}
z&=&\bar{t}_1-L(\bar{t}_1).
\label{t-F}
\end{eqnarray}
The formulas (\ref{ABG-par}), (\ref{ABG-impar}) and (\ref{ABF}) generalize those for $\tilde{A}$
and $\tilde{B}$ found in Ref. \cite{alves-granhen-silva-lima-2010}, which are valid
for a cavity with just the right boundary in movement.

From Eqs. (\ref{tensor}), (\ref{fGz}) and (\ref{fFz}), we get the exact formula for the renormalized energy density as
\begin{equation}
\left\langle T_{00}\left( t,x\right) \right\rangle =-f^{(s)}\left[\tilde{A}_{G}\left( v\right)+\tilde{A} _{F}\left( u\right)\right] -\tilde{B} _{G}\left( v\right) -\tilde{B} _{F}\left( u\right).
\label{tensor2}
\end{equation}
Eq. (\ref{tensor2}) gives directly the exact values 
for the energy density in a nonstatic cavity for 
arbitrary laws of motion $R(t)$ and $L(t)$.
Since $T_{00}=T_{11}$ in this model, we have the following exact formulas for the renormalized quantum forces ${\cal{F}}_{R}=\left\langle T_{00}\left[ t,R(t)\right] \right\rangle$ and ${\cal{F}}_{L}=-\left\langle T_{00}\left[ t,L(t)\right] \right\rangle$ 
(see Refs. \cite{alves-granhen-lima-2008,Alves-Farina-Maia-Neto-JPA-2003}) acting, respectively, on the right and left boundaries:
\begin{eqnarray}
{\cal{F}}_{R}(t) &=&-f^{(s)}\left\{ \tilde{A}_{G}\left[t+R(t)\right] +\tilde{A}_{F}\left[ t-R(t)\right] \right\} + \nonumber \\
&&-\tilde{B}_{G}\left[ t+R(t)\right] -\tilde{B}_{F}\left[ t-R(t)\right],
\label{forcaR}
\end{eqnarray}
\begin{eqnarray}
{\cal{F}}_{L}(t) &=&f^{(s)}\left\{ \tilde{A}_{G}\left[t+L(t)\right] +\tilde{A}_{F}\left[ t-L(t)\right] \right\} +  \nonumber \\
&&+\tilde{B}_{G}\left[ t+L(t)\right] +\tilde{B}_{F}\left[ t-L(t)\right].
\label{forcaL}
\end{eqnarray}
Next we examine the behavior of these forces in each region pointed in Fig. \ref{static-zones}.

In region I (Fig. \ref{static-zones}), we have $n_G=n_F=0$. Then,
Eqs. (\ref{ABG-par}) and (\ref{ABF}) give: 
$\tilde{A}_{G}\left( z\right)=\tilde{A}_{F}\left( z\right)=1$ and
$\tilde{B}_{G}\left( z\right)=\tilde{B}_{F}\left( z\right)=0$.
This results, as expected, in the static Casimir force 
$${\cal{F}}^{(Cas)}_{R}=-{\cal{F}}^{(Cas)}_{L}=-\pi/(24L_0^2),$$
acting on the boundaries.

In region II, we have $n_G=1$ and $n_F=0$.
For this case, Eq. (\ref{ABF}) gives
$\tilde{A}_{F}\left( u\right)=1$ and
$\tilde{B}_{F}\left( u\right)=0$,
whereas from Eq. (\ref{ABG-impar}) we have
$\tilde{A}_G(v)=A_R\left[t_1(v)\right]$ and
$\tilde{B}_G(v)=B_R\left[t_1(v)\right]$.
To calculate the force ${\cal{F}}_{R}(t) $ in Eq. (\ref{forcaR}) we do
$v\rightarrow t+R(t)$, and obtain
$t_1(v)$ as already discussed:
$t+R(t)=t_1+R(t_1)\Rightarrow t_1=t$.
Then we get $\tilde{A}_G\left[t+R\left(t\right)\right]=A_R\left(t\right)$ and
$\tilde{B}_G\left[t+R\left(t\right)\right]=B_R\left(t\right)$.
The force ${\cal{F}}_{R}(t)$ on the right boundary in 
region II, now relabeled as ${\cal{F}}_{R}^{(\mbox{\footnotesize II})}(t)$ is
\begin{equation}
\label{F-R-regiao-II}
{\cal{F}}_{R}^{(\mbox{\footnotesize II})}(t) =-f^{(s)}\left[1+A_R\left(t\right)\right]-B_R\left(t\right). 
\end{equation}
From this formula, we can obtain an analytical result for an arbitrary law of motion $R(t)$.
Note that in Eq. (\ref{F-R-regiao-II}) the subscript $L$ is not found, since the quantum force 
for the worldline in
region II has no influence of the movement of the left boundary.
Considering
the limit $L_0\rightarrow\infty$ we recover the quantum radiation
force ${\cal F}^{({\mbox{\footnotesize -u}})}_{q}$ corresponding
to the unbounded field, acting on the left side
of a single mirror:
$
\lim_{L_0\rightarrow\infty}{\cal{F}}_{R}^{(\mbox{\footnotesize II})}(t)=
{\cal F}^{({\mbox{\footnotesize -u}})}_{R}(t),
$
where
\begin{eqnarray}
{\cal F}^{({\mbox{\footnotesize -u}})}_{q}(t)&=&-B_{q}(t).
\label{unbounded-force-left}
\end{eqnarray}
In the non-relativistic limit, from (\ref{F-R-regiao-II}) we get 
${\cal{F}}_{R}^{(\mbox{\footnotesize II})}(t)\approx{\cal F}^{(Cas)}_{R}+{\dddot{R}}/{(12\pi)}$,
and adding the limit $L_0\rightarrow\infty$ we recover the approximate quantum radiation
force ${\cal{F}}_{R}^{(\mbox{\footnotesize II})}(t)\approx {\dddot{R}}/{(12\pi) }$, which acts on 
the left side of a single mirror \cite{Ford-Vilenkin-PRD-1982}.

In region III, we have $n_G=0$ and $n_F=1$.
For this case, Eqs. (\ref{ABG-par}) and (\ref{ABF}) 
give $\tilde{A}_{F}\left( u\right)=1/A_L\left[\bar{t}_1\left(u\right)\right]$; 
$\tilde{B}_{F}\left( u\right)=-B_L\left[\bar{t}_1\left(u\right)\right]/A_L
\left[\bar{t}_1\left(u\right)\right]$; $\tilde{A}_G(v)=1$; $\tilde{B}_G(v)=0$. 
Considering $u\rightarrow t-L(t)$ and $t-L(t)=\bar{t}_1-L(\bar{t}_1)\Rightarrow \bar{t}_1=t$,
the force ${\cal{F}}_{L}(t)$ on the left boundary in this
region, now relabeled as ${\cal{F}}_{L}^{(\mbox{\footnotesize III})}(t)$ is
\begin{equation}
{\cal{F}}_{L}^{(\mbox{\footnotesize III})}(t) =f^{(s)}\left\{ 1 +\frac{1}{A_L\left(t\right)}
 \right\}-\frac{B_L\left(t\right)}{A_L\left(t\right)}.
\label{F-L-regiao-III}
\end{equation}%
Considering
the limit $L_0\rightarrow\infty$ we recover the quantum radiation
force ${\cal F}^{({\mbox{\footnotesize +u}})}_{q}$ corresponding
to the unbounded field, acting on the right side
of a single mirror:
$
\lim_{L_0\rightarrow\infty}{\cal{F}}_{L}^{(\mbox{\footnotesize III})}(t)=
{\cal F}^{({\mbox{\footnotesize  +u}})}_{L}(t),
$
where
\begin{eqnarray}
{\cal F}^{({\mbox{\footnotesize  +u}})}_{q}(t)&=&-\frac{B_q\left(t\right)}{A_q\left(t\right)}.
\label{unbounded-force-right}
\end{eqnarray}
From Eqs. (\ref{unbounded-force-left}) and  (\ref{unbounded-force-right}) 
we recover the total quantum force ${\cal F}^{({\mbox{\footnotesize u}})}_{q}(t)$ 
acting on a single mirror at vacuum, with a prescribed trajectory $x=q(t)$:
\begin{eqnarray}
{\cal F}^{({\mbox{\footnotesize u}})}_{q}(t)&=&{\cal F}^{({\mbox{\footnotesize  -u}})}_{q}(t)+
{\cal F}^{({\mbox{\footnotesize  +u}})}_{q}(t)
\nonumber
\\
&=&\left( 1+\dot{q}^{2}\right)
\left\{[{\ddot{q}^{2}\dot{q}}/{(2\pi) }]
/{\left( 1-\dot{q}^{2}\right) ^{4}}\right.
\nonumber\\
&&\left.+[{
\dddot{q}}/{(6\pi) }]/{\left( 1-\dot{q}
^{2}\right) ^{3}}\right\}\nonumber,
\end{eqnarray}
which is in agreement with that found in literature
(see Ref. \cite{alves-granhen-lima-2008}).
In the non-relativistic limit, we reobtain the approximate quantum radiation
force 
${\cal F}^{({\mbox{\footnotesize u}})}_{q}(t)\approx{\dddot{q}}/{(6\pi)}$ \cite{Ford-Vilenkin-PRD-1982}.

To compute the total forces ${\cal{F}}^{(tot)}_{R}$ and ${\cal{F}}^{(tot)}_{L}$
acting on, respectively, the right and left boundaries, for any of the regions II, III or IV
showed in Fig. \ref{static-zones}, we need, in addition
to Eqs. (\ref{forcaR}) and (\ref{forcaL}), to take into account
the remaining dynamical Casimir forces corresponding to the vacuum field outside the cavity,
which are given by Eqs. (\ref{unbounded-force-left}) and (\ref{unbounded-force-right}). We write:
\begin{eqnarray}
{\cal{F}}^{(tot)}_{R} &=&{\cal{F}}_{R}(t)+ {\cal F}^{({\mbox{\footnotesize  +u}})}_{R}(t), 
\label{forcaR-total}
\\
{\cal{F}}^{(tot)}_{L} &=&{\cal{F}}_{L}(t)+ {\cal F}^{({\mbox{\footnotesize  -u}})}_{L}(t). 
\label{forcaL-total}
\end{eqnarray}
Eqs. (\ref{forcaR-total}) and (\ref{forcaL-total}) enable us to calculate directly 
and analytically the total quantum forces acting
on both mirrors for arbitrary laws of motion $R(t)$ and $L(t)$, in regions II or III,
because for these regions Eqs. (\ref{forcaR}) and (\ref{forcaL}) are replaced
by their particular cases given by Eqs. (\ref{F-R-regiao-II}) and (\ref{F-L-regiao-III}).

In region IV (see Fig. \ref{static-zones}), 
in general it is difficult to obtain exact analytical results 
for the quantum forces (\ref{forcaR}) and (\ref{forcaL}), for  
arbitrary trajectories $R(t)$ and $L(t)$. 
The difficulty is in solving equations like 
$t_1-R\left(t_1\right)=t_2-L\left(t_2\right)$
(see Eq. (\ref{t-G})),
which arise after a second reflection ($n_G\geq 2$ or/and $n_F\geq 2$).
Trajectories can be constructed to give analytical solutions to these equations,
but a large class of relevant laws of motion do not result in exact analytical solutions.
However, our results enable us
to obtain exact numerical results for the quantum force acting on the moving boundaries of a
cavity for an arbitrary law of movement, including non-oscillating movements with large amplitudes,
which are out of reach of the perturbative approaches found in the literature, as we will examine next. 
In this context, let us apply our formulas to the following particular non-trivial trajectory, which is based on the
one proposed by Haro in Ref. \cite{Haro-JPA-2005}:
\begin{subequations}
\label{haros}
\begin{eqnarray}
L\left(t\right)&=& \kappa_L \ln [\cosh (t)], \label{LHaro} \\
R\left(t\right)&=& L_{0}+\kappa_R \ln [\cosh (t)]. \label{RHaro}
\end{eqnarray}
\end{subequations}
\begin{figure}
\subfigure[\label{trajectories}]{\includegraphics[scale=.45]{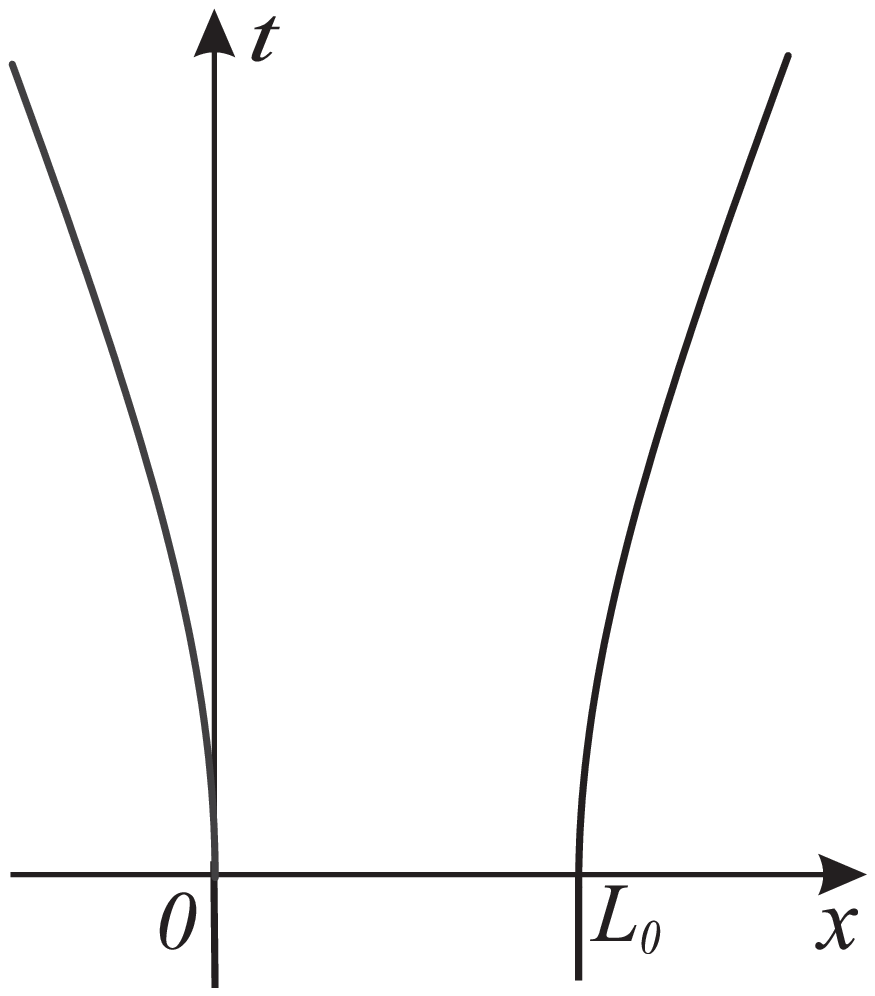}}
\qquad
\subfigure[\label{trajectories-2}]{\includegraphics[scale=.45]{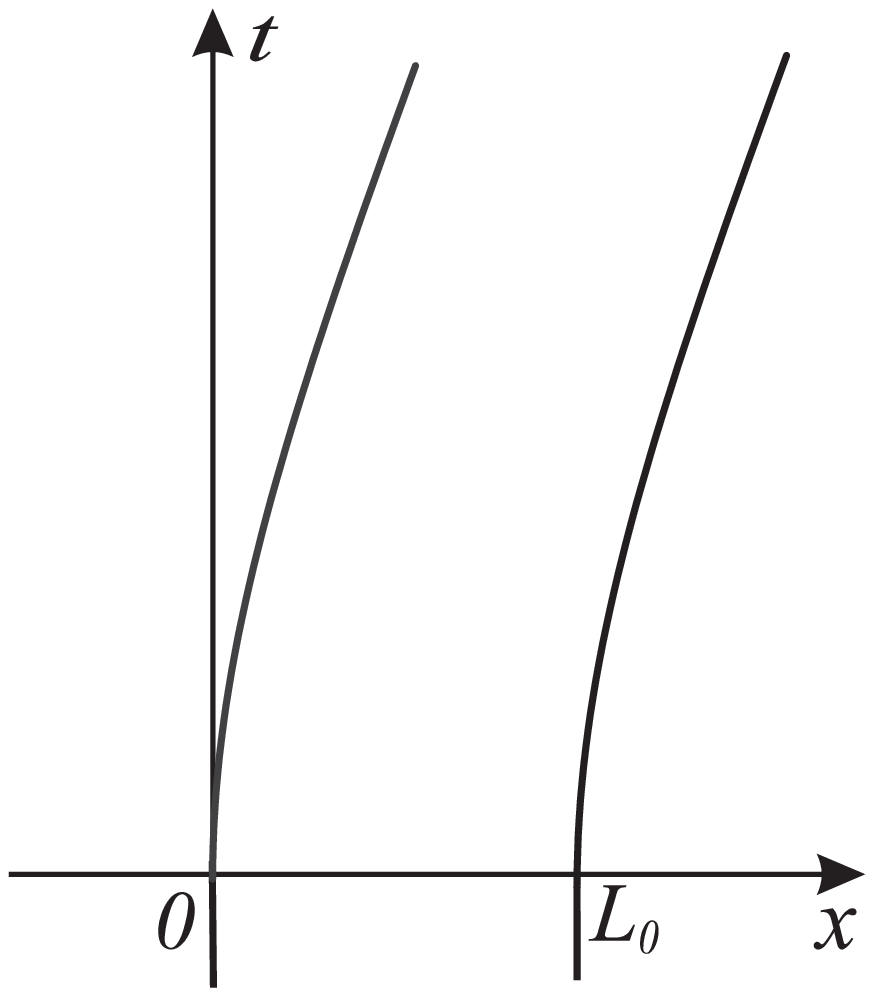}}
\caption{The solid lines show the boundaries trajectories described in Eq. (\ref{haros}).
Fig. \ref{trajectories} describes the case $\kappa_R=-\kappa_L=0.1$, whereas \ref{trajectories-2} describes
the case $\kappa_R=\kappa_L=0.1$.}
\end{figure}
Considering, for instance, $\kappa_R=-\kappa_L=0.1$ (Fig. \ref{trajectories}), we have an expanding cavity
with large amplitude and
relativistic velocities. If we consider $\kappa_R=\kappa_L=0.1$ (Fig. \ref{trajectories-2}), we have
the mirrors in movement with relativistic velocities, but keeping constant the cavity length. 

\begin{figure}
\includegraphics[scale=.35,angle=0]{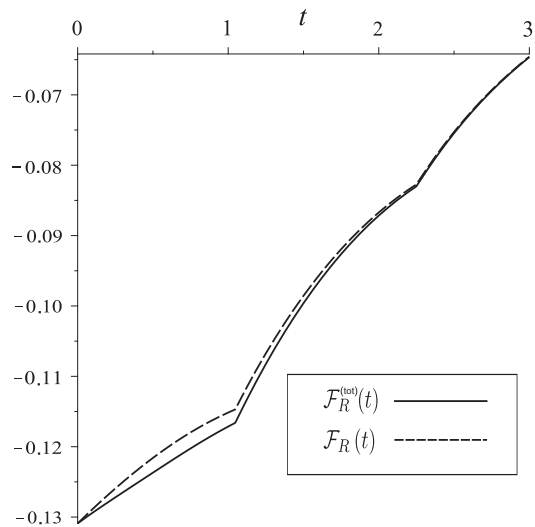}
\caption{The solid line shows the total force ${\cal{F}}^{(tot)}_{R}(t)$, 
whereas the dashed line shows the force ${\cal{F}}_{R}(t)$,
both for the law of movement (\ref{haros}), 
with $\kappa_R=-\kappa_L=0.1$ and $L_0=1$.}
\label{breathing-case}
\end{figure}

\begin{figure}
\includegraphics[scale=.35,angle=0]{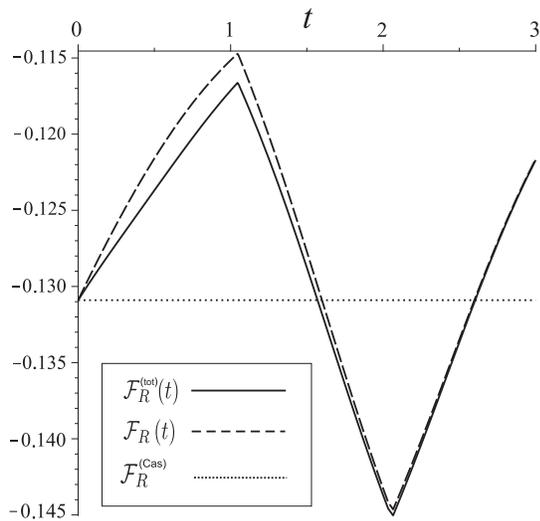}
\caption{The solid line shows the total force ${\cal{F}}^{(tot)}_{R}(t)$, 
the dashed line shows the force ${\cal{F}}_{R}(t)$,
both for the law of movement (\ref{haros}), 
with $\kappa_R=\kappa_L=0.1$ and $L_0=1$. The dotted line shows the static Casimir force
.}
\label{translational-case}
\end{figure}

In Figs. \ref{breathing-case} and \ref{translational-case}, 
using our formulas (\ref{ABG-par})-(\ref{ABF}) and (\ref{forcaR-total}),
we plot the time evolution of the 
quantum force ${\cal{F}}^{(tot)}_{R}(t)$ and ${\cal{F}}_R\left(t\right)$,
for, respectively, the cases
$\kappa_R=-\kappa_L=0.1$ (see Fig. \ref{trajectories}), and
$\kappa_R=\kappa_L=0.1$ (see Fig. \ref{trajectories-2}).
We can see discontinuities of the derivatives for ${\cal{F}}^{(tot)}_{R}$ and ${\cal{F}}_R\left(t\right)$.
These discontinuities always occur when the front of
the wave in the energy density meets the right boundary.
In the case, for instance, showed in Fig. \ref{breathing-case},
when $t=0$ the left boundary starts to move and generate a wave in the energy density,
propagating rightward and meeting the right boundary at the instant $t={\tau}_1 \approx 1.05$,
calculated via equation ${\tau}_1-R\left({\tau}_1\right)=0$, and 
which corresponds to the first discontinuity of the derivative showed in Fig. \ref{breathing-case}. 
At $t=0$, another front of wave is generated by the right boundary, propagating leftward and meeting the left boundary at the instant $\tau_1 \approx 1.05$, and then reflected back and meeting the right boundary at the instant $\tau_2 \approx 2.25$, calculated from the equation $\tau_2-R\left(\tau_2\right)=\tau_1-L\left(\tau_1\right)$. 
This instant corresponds to the second discontinuity of the derivative showed in  Fig. \ref{breathing-case}. 
Since the length of the cavity remains the same in the case
showed in Fig. \ref{trajectories-2},
the quantum force ${\cal{F}}^{(tot)}_{R}(t)$ oscillates around
the static Casimir force (Fig. \ref{translational-case}), whereas it goes to zero in the case
showed in Fig. (Fig. \ref{breathing-case}) where the boundaries
go to an asymptotic behavior of infinity lenght and constant velocity.

Summarizing our results, the formulas obtained in the present paper enable us to get directly exact values 
of the energy density of the field and the quantum force acting on the boundaries
of a nonstatic cavity for 
arbitrary laws of motion for the moving boundaries, for vacuum as the initial state of the field. 
Eqs. (\ref{fGdeR}) and (\ref{fGdeL}) are an extension of the corresponding
equation for a cavity with just one moving boundary found in Ref. \cite{Cole-Schieve-2001},
and the achievement of $f_G$ and $f_F$ recursively, tracing back a sequence of null lines,
can be viewed as an extension of the work done in Ref. \cite{Li-Li-PLA-2002}. 
Formulas (\ref{ABG-par}), (\ref{ABG-impar}) and (\ref{ABF}) generalize those 
found in Ref. \cite{alves-granhen-silva-lima-2010}.
For the particular cases of a cavity with just one moving boundary, 
non-relativistic velocities, or in the limit of infinity length of the cavity
(a single mirror), our results are in agreement with those found in the
literature \cite{Fulling-Davies-PRS-1976-I, Ford-Vilenkin-PRD-1982,Cole-Schieve-2001, alves-granhen-lima-2008,alves-granhen-silva-lima-2010}.

The present results enable investigation of several problems
(usually treated by perturbative approaches in the literature)
with an exact approach and also out of the regime of small amplitudes. For instance,
those related to the inertial forces in
the Casimir effect with two moving mirrors \cite{Machado-Maia-Neto-PRD-2002}, 
or the interference phenomena in the photon production \cite{Ji-Jung-Soh-1998}.
These issues are under investigation and will be discussed in future papers.

We acknowledge  A. L. C. Rego, 
C. Farina and P. A. Maia Neto 
for valuable discussions. 
We are grateful to C. Farina, A. L. C. Rego and H. O. Silva for careful reading 
of this paper. This work was supported by CNPq and CAPES - Brazil.


\end{document}